# Electromagnetic Properties of Topological Crystalline Superconductor $Sn_{0.5}In_{0.5}Te$

V. K. Maurya, Shruti, P. Srivastava and S. Patnaik[*]

*School of Physical Sciences, Jawaharlal Nehru University New Delhi, India*
*\*Email: spatnaik@mail.jnu.ac.in*

We report on the superconducting properties of indium doped SnTe which has recently been explored as a topological crystalline superconductor. Single crystals of $Sn_{0.5}In_{0.5}Te$ have been synthesized by modified Bridgman method. Resistivity measurement performed in the range 1.6 K to 300 K shows metallic normal state with onset of superconducting transition at $T_c$ = 4.5 K. Bulk superconductivity has also been confirmed by DC magnetization, AC susceptibility and rf penetration depth measurements. The zero temperature upper critical field, lower critical field, coherence length, and penetration depth are estimated to be 1.6 T, 10 Oe, 143.5 Å and 832 nm respectively. Temperature dependence of low temperature penetration depth indicates S-wave fully gapped characteristics with BCS gap $\Delta_0 = 1.247$ meV. Hall and Seebeck coefficient measurements confirm dominance of hole conduction with possible phonon-drag effects around ~45 K. Resistive transition studied under applied magnetic field shows thermally activated flux flow behavior.

## Introduction

The discovery of superconductivity in degenerate many-valley semiconductor SnTe traces its origin to its theoretical prediction by Cohen [1] and its subsequent experimental verification by R. A. Hein [2]. Way back in 1966, the dependence of the critical magnetic fields and Ginzburg –Landau parameter of SnTe on carrier concentration (due to Sn vacancies) was reported. The superconducting transition temperature $T_c$ was found to increase with increasing carrier concentration from 0.034 K ($7.5\times10^{20}$ per cm$^3$) to 0.214 K ($20.0\times10^{20}$ per cm$^3$). The relevance of quasi-local impurity states achieved through In/Tl doping onto SnTe/PbTe leading to superconductivity were also studied [3]. In a revival of sort, around 2010, there were reports of remarkable increase in $T_c$ to ~2 K when SnTe was doped at the Sn site with In. Conspicuously, this was achieved for similar carrier concentration as compared to the off-stoichiometric specimen [4]. Very recently, this novel system of $Sn_{1-x}In_xTe$ has been assigned to an exotic class of Topological Crystalline Superconductor (TCS) and optimal superconducting $T_c$ of 4.7 K [5] and 4.5 K [6] has been reported.

The research of 1960s identified that superconductivity appeared at a carrier concentration corresponding to a second valence band filling in SnTe; a phenomena that was quickly generalized to other IV – VI semiconductors. Hein et al. [7] confirmed that it was indeed inter-valley scattering which was important in the determination of the BCS electron-phonon coupling parameter *V*. If interband or intervalley scattering is possible, then that can help overcome the Coulomb repulsion leading to superconductivity. But there have been suggestions that the manifold increase in transition temperature with In doping could possibly be not due to a mechanism associated with inter-valley scattering. Notably, indium is known to skip the +2 valence, in favour of +1 and +3 in most compounds. This tendency leads to the suggestion as to whether valance fluctuations associated with In impurities in SnTe might play a dominant role in the enhanced superconductivity [4].

In the recent past, the renewed interest in SnTe relates to its characterization as a topological crystalline insulator (TCI) [8]. A topological insulator (TI) is an unusual quantum state of matter that is protected by time-reversal symmetry and possesses a full band gap in the bulk but has gapless surface states. The discovery of TI prompted the search for other topological states which may be protected by other symmetries and one such candidate is the TCI in which the metallic surface states are protected by the mirror symmetry of the crystal structure. Such TCI phase in SnTe has been confirmed by Tanaka et al. by angle-resolved photoemission spectroscopy (ARPES) [9]. The characteristic metallic Dirac-cone surface band has been confirmed. Curiously, such gapless surface states have not seen in PbTe, but the narrow-band-gap semiconductor $Pb_{0.77}Sn_{0.23}Se$ undergoes a phase transition from TI to TCI phase as a function of temperature [10]. Very recently, Topological Crystalline Superconductivity (TCS) in $Sn_{1-x}In_xTe$ (x=0.045) has been confirmed through the existence of surface Andreev bound state by performing point-contact spectroscopy [11]. This is apparently a hallmark of unconventional superconductivity. Thin film of the topological crystalline insulator SnTe has also been grown epitaxially on $Bi_2Te_3$ that has made it possible for the observation of the topological surface state in a TI [12].

The critical question that remains to be answered is whether the superconductivity in $Sn_{1-x}In_xTe$, derived from a TCI phase, shows any departure from BCS superconductivity as evidenced in Cu doped topological insulator $Bi_2Se_3$ [13]. In this paper, we report on growth and extensive characterization of single crystals $Sn_{0.5}In_{0.5}Te$ (with optimal $T_c$) and try to ascertain its basic superconducting properties and identify in what way it may be different from a normal impurity scattering driven degenerate semiconductor-superconductor. Specifically, we report on the successful synthesis of 50% In doped SnTe (with optimal $T_c$) single crystals that indicate signatures of S-wave symmetry with robust type–II characteristics. We also estimate the important superconducting parameters such as the critical fields, coherence length, penetration depth and the scaling parameters for the observed thermally activated flux flow.

**Experimental Methods**

Single crystals of $Sn_{1-x}In_xTe$ were prepared by modified Bridgman method for various doping (x = 0, 0.25, 0.4 and 0.5), but in this paper, we study the properties of the specimen with optimal $T_c$ (x = 0.5). Single crystals were obtained by melting stoichiometric amounts of high purity elemental powder of Sn (99.99%), Te (99.999%) and shots of In (99.999%) at 900°C for 5 days in sealed evacuated quartz tubes. Intermittent shaking was performed for the homogeneity of melt sample. Sample was slowly cooled to 770°C over a period of 72 hours, then annealed at 770°C for 48 hours followed by fast cooling to room temperature. Silvery-shiny single crystals were cleaved along z-axis. X-ray Diffraction was carried out on the powdered samples by *RIGAKU* powder X- ray Diffractometer (Miniflex 600). Resistivity, Hall, and magnetization measurements were performed on a *Cryogenic* Physical Properties Measurements System (PPMS). RF Penetration depth was measured on a separate 8 Tesla Cryogen Free Magnet in the temperature range 1.6 K to 6 K. Thermo-electric power was measured in bridge geometry across two copper blocks in conjunction with chip heaters to produce a dynamic temperature gradient.

## Results and Discussion

The XRD pattern of $Sn_{0.5}In_{0.5}Te$ presented in Figure 1 is in accordance with the reference data from JCPDF (no. 089-3974) that confirm phase purity of the as-grown samples. The specimen crystallizes in rock-salt structure with space group $Fm\bar{3}m$. Rietveld refinement of observed XRD peaks was carried out with the help of *FULLPROF* software. The $\chi^2$ is found to be 1.76, which indicates a good fitting to the experimental data. In the inset (a) of Figure 1, the schematic representation of the cubic crystal structure is depicted. The calculated lattice parameters are **a** = **b** = **c** = 6.265 Å and cell volume is 245.65 Å$^3$. In the inset 1(b), we show the surface morphology images obtained from a *Zeiss EVO40* SEM analyzer. From SEM image it is clear that the crystals were cleaved in clear planes and we can see clear and smooth surface at the 20 μm magnification-range. Figure 1(c) shows EDAX spectrum data (*Bruker AXS Microanalysis*) that confirms intended atomic percentage doping of indium in as-grown samples.

Resistivity data for $Sn_{0.5}In_{0.5}Te$ measured by four probe method in the temperature range 1.6 K to 300 K are shown in Figure 2. The compound shows robust metallic behavior in the normal state. As shown in the inset (a) of Figure 2 (in an expanded temperature scale), a superconducting transition is observed with $T_c^{onset}$ at 4.5 K and $T_c^{zero}$ at 4.3 K. The resistivity changes from 0.88 mOhm-cm at 300K to 0.67 mOhm-cm just above the transition temperature. The corresponding residual resistivity ratio (RRR) is found to be 1.31. Inset (b) of Figure 2 shows the resistive transitions ρ(T) in the presence of external magnetic fields applied perpendicular to *ab* plane of the sample as well as to the transport current direction. The transition width between $T_c^{onset}$ and $T_c^{offset}$ is about 0.2 K in low magnetic fields but that increases up to ~ 0.6 K for higher fields.

Inset (c) of Figure 2 shows upper critical field $H_{c2}$ and irreversibility field $H_{irr}$ as estimated from the magneto-resistance data from Figure 2(b). The criteria used from defining $H_{c2}$ (corresponding to complete suppression of superconductivity) and irreversibility field $H_{irr}$ (corresponding to onset of dissipation) is schematically defined in Figure 2(b). The numerical value of $H_{c2}(0)$ was calculated by applying the Werthamer-Helfand-Hohenberg (WHH) formula: $H_{c2}(0) = -0.69T_c(dH_{c2}/dT|_{Tc})$. Using the value of slope $dH_{c2}/dT|_{Tc}= -0.496$, the value of $H_{c2}(0)$ is found to be 1.6 Tesla. Similar measurements for field applied parallel to crystal plane provided identical slope (and $H_{c2}(0)$) confirming isotropic behavior of superconducting parameters. From the above value of $H_{c2}(0)$, Ginzburg-Landau coherence length $\xi = (2.07*10^{-7}/2\pi H_{c2})^{1/2}$ is estimated to be $\xi_{GL}(0) \sim 143.5$ Å. For the determination of lower critical field ($H_{c1}$), we have used the deviation from linearity method (in Meissner state) from magnetization versus magnetic field measurements. The value of $H_{c1}$ estimated from iso-thermal magnetization measurements at low temperature was ~ 10 G (data not shown). Further, $H_{c1}$ is related to value of penetration depth and coherence length as follows [14].

$$H_{c1} = \frac{\Phi_0}{4\pi\lambda^2}(\frac{\ln \lambda}{\xi} + 0.485)$$  [1]

Using $H_{c1}=10$ G, $\xi_{GL}(0)\sim 143.5$ Å and $\Phi_0 =2.0\times10^{-7}$ G-cm$^2$, the zero temperature value of penetration depth was found to be 853 nm. Hence the value of Ginzburg-Landau parameter κ (= λ/ξ) is estimated to be 59.4. This puts this compound in the category of extreme type-II superconductivity.

Figure 3 shows the variation of DC magnetization with temperature for $Sn_{0.5}In_{0.5}Te$ under ZFC (Zero Field Cooled) and FC (Field Cooled) protocols at an applied magnetic field of 10 Oe. The $T_c$ for this sample can be observed around 4.4 K that is slightly lower than what is observed in the resistivity measurement. Isothermal MH loop at constant temperature T= 2 K in fields up to 2 T is shown in the inset Figure 3(a). The corresponding nominal critical current density $J_c$ of the sample with rectangular cross-section (a × b with a ~ 2.76 mm , b~ 2.86 mm) perpendicular to the applied field, was calculated from the Bean Critical State expression $J_c = (20 \Delta m)/(V a(1 - a/3b))$, where $V$ is the sample volume, $\Delta m = m^- - m^+$, and $m^+(m^-)$ is the moment associated with increasing (decreasing) field. The critical current density $J_c$ versus magnetic field at T = 2 K was derived and it was found that the $J_c$ decreases dramatically from the zero field value $3.4 \times 10^3$ A cm$^{-2}$ to 60.5 A cm$^{-2}$ at H= 0.5 T. The AC susceptibility measured in an AC drive field with a frequency of 781 Hz is shown in inset 3(b). A DC applied field was superimposed parallel to the AC field (~5 Oe) to check the AC losses in the mixed state. The real part ($\chi'$) of the AC susceptibility showed a sharp transition to diamagnetism at ~4.4 K (with zero DC field) confirming onset of superconductivity in Sn$_{0.5}$In$_{0.5}$Te. At a DC field of 2.5 kOe, as seen in inset 3b, $\chi'$ showed rapid suppression of the superconductivity due to dominant vortex dynamics.

The temperature dependence of London penetration depth gives important information about the pairing symmetry and superconducting gap [15]. Using an ultra-stable tunnel diode oscillator technique, in Figure 4 we plot the change in penetration depth $\Delta\lambda$ as a function of reduced temperature (T/T$_c$). Inset (a) of Figure 4 shows the change in penetration depth $\Delta\lambda$ scaled with $\lambda_0$ (total change in penetration depth) as a function of temperature measured down to the lowest temperature of 1.69 K. A sharp change in data at T$_c$ = 4.4 K reconfirms the onset of bulk diamagnetic state. The measured quantity, shift in the resonant frequency, $\Delta F \equiv f_s - f_0$ is proportional to $\chi(T)$, where $\chi$ is the total magnetic susceptibility, $f_0$ is the resonance frequency in absence of sample, and $f_s$ is the resonance frequency in presence of sample. The change in penetration depth is defined as $\Delta\lambda(T) = -G\delta F(T)$ where $\delta F(T) = \Delta F(T) - \Delta F(T)_{min}$ and G =R $[2V_c (1 - N)]/ f_0 V_s$ is a geometric calibration factor defined by the coil ($V_c$), and the sample ($V_s$) volumes and the demagnetization factor N [16]. We have calibrated our system for pure niobium sample of similar shape and size and G comes out to be 2.27 for our apparatus.

The main panel of Figure 4 also shows the fitting of the data to conventional BCS model. According to BCS model, for a fully gapped S-wave superconductor, the penetration depth at low temperature is given by [17]

$$\Delta\lambda(T) = \lambda(0)\sqrt{\frac{\pi\Delta_0}{2k_BT}}\exp\left(\frac{-\Delta_0}{k_BT}\right) \quad [2]$$

Where $\Delta_0$ and $\lambda(0)$ are the zero temperature values of energy gap and penetration depth respectively. For a d-wave pairing on the other hand [18],

$$\Delta\lambda(T) = \lambda(0)\frac{2\ln 2}{\alpha\Delta_0}T \quad [3]$$

In special cases, impurity scattering changes this temperature dependence of $\Delta\lambda$ from T to $T^2$ [19]. The $\Delta\lambda$ data of Figure 4 are very well fitted to Eq. 2 up to $T/T_c \leq 0.5$ as shown by red solid line. The gap value found from fitting was $\Delta_0 = 1.247$ meV with corresponding gap ratio $2\Delta_0/k_BT_c = 6.4$ and $\lambda(0) = 832$ nm which is very close to calculated value of 853 nm from Equation 1. The value of gap ratio is higher than weak coupling BCS value of 3.53 which suggests strong coupling features in $Sn_{0.5}In_{0.5}Te$. This is in contrast to the weak-intermediate coupling projection from specific heat measurements on $Sn_{0.6}In_{0.4}Te$ [5]. The penetration depth flattens at very low temperature values indicating fully developed superconducting gap in this compound. Node-less superconducting gap has also been observed in thermal conductivity measurement of $Sn_{0.6}In_{0.4}Te$ [20]. On fitting the low temperature data with power law $T^n$ gives n ~6 which rules out the possibility of d-wave pairing in $Sn_{0.5}In_{0.5}Te$.

The Seebeck coefficient (S) of single crystals of $Sn_{0.5}In_{0.5}Te$ with respect to copper is shown in inset 4(b). S is positive in the whole temperature range as verified from Hall measurements as well. At the lowest measured temperature (12K), S has a maximum value of 2.24 µV / K but it curiously shows a peak structure at ~45 K with varying temperature. While the origin of such sharp peak needs more detailed analysis, for the present it would suffice to report that such peaks in the temperature dependence of thermopower (TEP) is usually observed due to the phonon-drag effects. The Hall coefficient ($R_H$) did not show any sharp variation around this temperature and the positive value of 0.09 $cm^3$/C remained almost constant between 12K and 100K (data not shown). No such peak structure was observed in Seebeck coefficient of parent compound SnTe either, although the magnitude of Seebeck coefficient was expectedly higher.

Under the gamut of type-II superconductivity, the broadening of the resistive transition, $\rho(T) < 1\% \rho_n$ (where $\rho_n$ is the resistivity in the normal state just above the transition), in a magnetic field is interpreted in terms of a dissipation of energy caused by the motion of vortices. This interpretation is based on the fact that for the low-resistance region, the resistance is caused by the creep and flow of magnetic vortices. In the mixed state, the flux lines will be pinned due to various interactions, e.g., impurities, stress, extended defects, etc. During flux creep, flux line or flux bundles can be thermally activated over the pinning energy barrier, even if the Lorentz force exerted on the flux bundle by the current is smaller than the pinning force. The $\rho(T)$ dependences of such thermally activated origin are described by the Arrhenius equation $\rho = \rho_0 \exp(-U_0/k_B T)$, where, $U_0$ is the flux-flow activation energy. This can be obtained from the slope of the linear parts of an Arrhenius plot and $\rho_0$ is a field-independent pre-exponential factor. Investigations of high-$Tc$ superconductors and artificial multilayers have showed that the activation energy exhibits different power-law dependences on a magnetic field, i.e. $U_0(B) \sim B^{-n}$. Figure 5 shows plots of ln ρ vs. 1/T for the calculation of activation energy $U_0$ at various fields from 0 Tesla to 0.9 Tesla. The activation energy $U_0$ is determined from the slope of the curve in this linear region. We get the straight lines over the 5 decades of the resistivity which validates the thermally activated flux flow (TAFF) defined by Arrhenius law. The activation energy varies from $U_0/k_B \sim 500.4$ K to 142.5 K for the magnetic field of H = 0.05 T and H = 0.9 T, respectively. The magnetic field versus the activation energy $U_0$ plot shown in the Inset 5(a) suggests the power law dependence on magnetic field $U_0$, with n = 0.6 (<0.9 T) A rapid decrease of the activation energy for the sample in region $B > 0.9$ T was also observed that reflected a dramatic loss of the current carrying capabilities of the superconductor due to the weakening of the meagre flux line pinning with increasing magnetic field. This was also confirmed from the remanent magnetization experiment results of which are summarized in inset 5(b). In this experiment an external field was applied and then removed followed by measurement of the remanent magnetization. We observe a sharp single peak in $dm_R/d(\log\mu_0 H)$ versus $\log(\mu_0 H)$ at 300 Oe which indicates only one dominant scale of current loops. This implies that only intragrain super current exists and intergrain pinning is negligible.

## Conclusions

In conclusion, we have successfully synthesized single crystals of $Sn_xIn_{1-x}Te$ and an optimal resistive $T_c$ onset of 4.5 K is confirmed for x= 0.5. Calculated value of $H_{c2}(0)$ is found to be 1.6 Tesla under WHH extrapolation. The corresponding Ginzburg-Landau coherence length is estimated to be 143.5 Å. The zero temperature London penetration depth is estimated to be 832 nm. Low temperature penetration depth fits well to S-wave pairing symmetry and no strong deviation from isotropic BCS theory is observed. In essence, no unconventional superconductivity as expected for a superconductor derived from a topological crystalline insulating phase is observed.

# Acknowledgement

V. K. Maurya, Shruti, and P. Srivastava acknowledge the UGC (India) for providing UGC-BSR, UGC-SRF, and UGC-DS Kothari fellowships respectively. Technical support from AIRF (JNU) is gratefully acknowledged. SP would acknowledge financial support from DST- PURSE and DST –FIST programs of Department of Science and Technology (Government of India).

# Figure captions

**Figure 1.** Rietveld refinement of observed powder XRD peaks fitted with the calculated peaks for single crystal $Sn_{0.5}In_{0.5}Te$. The $\chi^2$ is found to be 1.76. In the inset (a) crystal structure is represented schematically. In inset (b) SEM image shows the fine and smooth surface cleaved for $Sn_{0.5}In_{0.5}Te$ crystal at 20μ magnification. Inset (c) shows EDAX spectrum.

**Figure 2.** Resistivity as a function of temperature is plotted for $Sn_{0.5}In_{0.5}Te$ crystal from 1.6 K to 300 K. In a broadened scale, superconducting transition at 4.5 K is indicated in inset 2(a) In the inset 2(b) resistivity is plotted as a function of temperature in the presence of DC magnetic fields (0 T to 1.2 T in constant field in the step of 0.1 T). Inset 2(c) shows the upper critical field $H_{c2}$ and the irreversibility field $H_{irr}$ as a function of temperature. Field was applied perpendicular to the crystal plane and the direction of transport current.

**Figure 3.** DC magnetization measurement under ZFC/FC protocol is shown for an external field of 10 Oe. The data have been taken during the warming cycle. A typical M-H curve at T = 2 K is shown in the inset (a). Real part of AC susceptibility is shown in the inset (b) for external DC field of 0 T and 0.25 T.

**Figure 4.** Change in penetration depth as a function of reduced temperature is plotted. The red solid curve shows the fitting using BCS Equation 2. Inset (a) shows the change in penetration depth normalized to the total change in penetration depth down up to 1.69 K. Temperature dependence of Seebeck coefficient for $Sn_{0.5}In_{0.5}Te$ is shown in inset (b). The Seebeck coefficient is positive in the whole temperature range.

**Figure 5.** Arrhenius plot of the resistivity for $Sn_{0.5}In_{0.5}Te$ for (left to right) with H = 0 to 0.9 T. Dependence of the activation energy $U_0/k_B$ on magnetic field for the superconducting sample is shown in inset (a). In the inset (b), $dm_R/d(\log\mu_0H)$ versus $\log(\mu_0H)$ is plotted which shows a sharp single peak at ~300 Oe indicating absence of intergrain pinning.

Figure 1

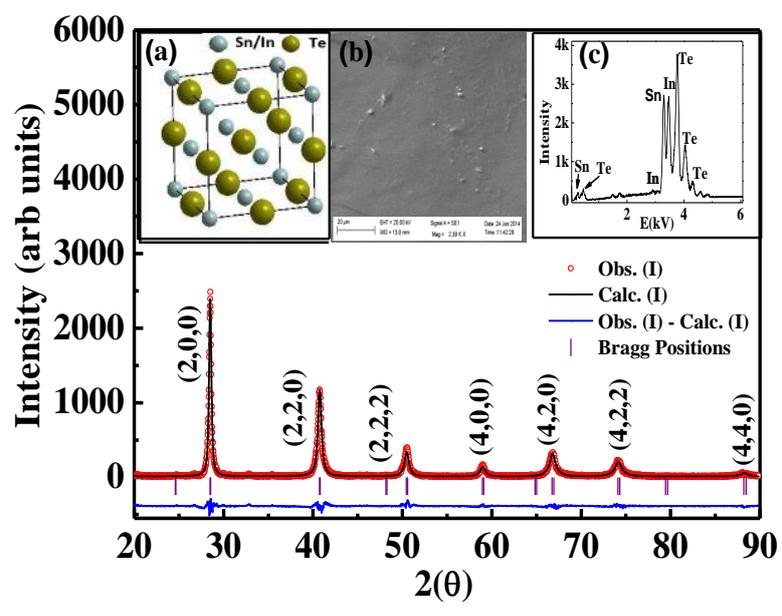

Figure 2

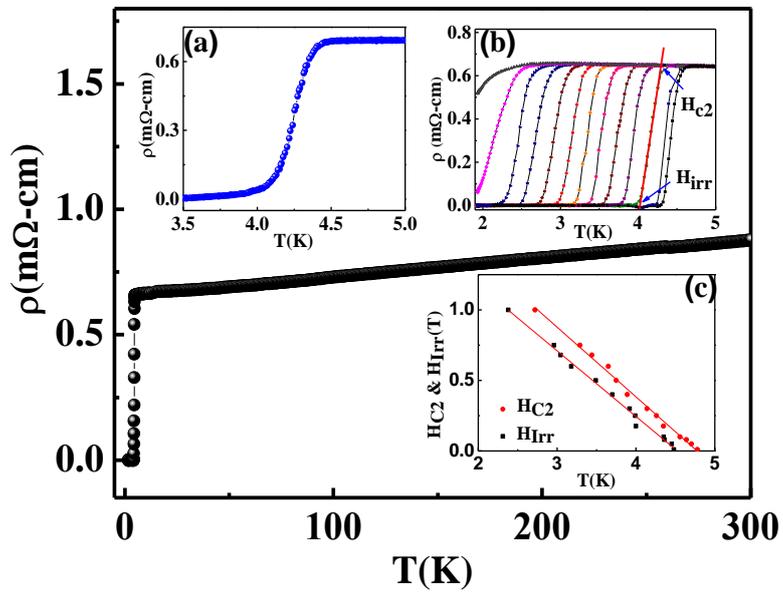

Figure 3

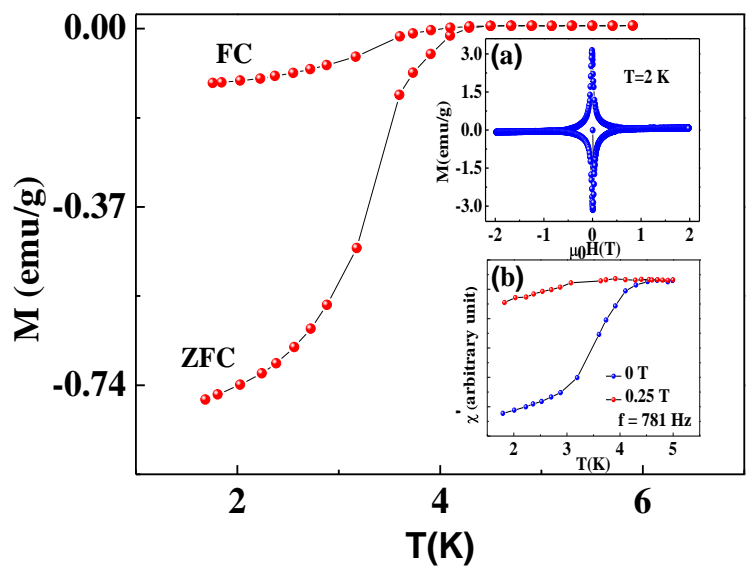

**Figure 4**

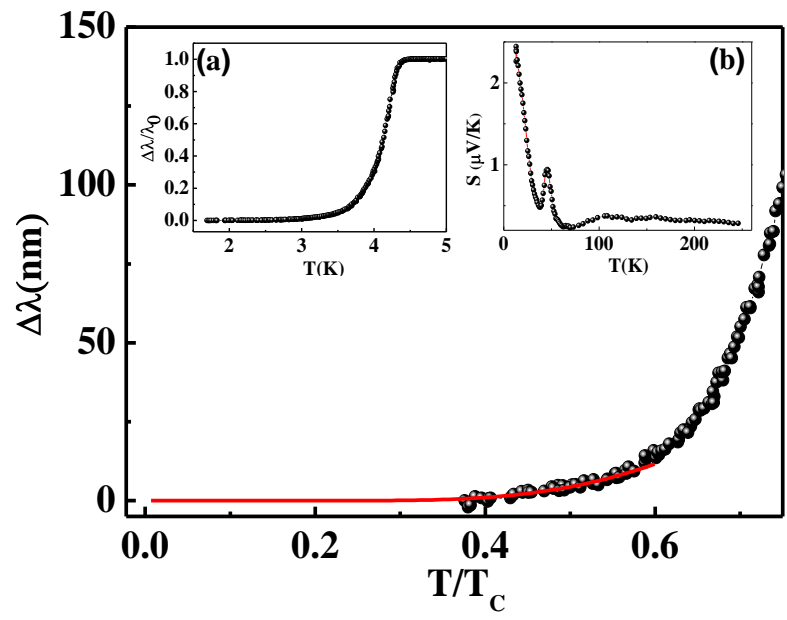

**Figure 5**

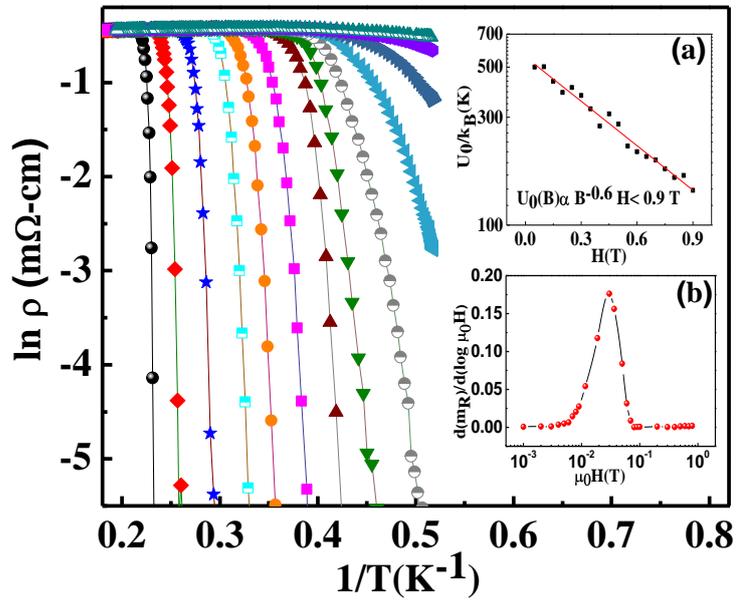